\documentstyle[11pt,preprint,psfig]{aastex}
\def\mearth{{\rm\,M_\oplus}}
 
\begin{document}
 
\title{Making other Earths: Dynamical Simulations of Terrestrial Planet
  Formation and Water Delivery}
 
\author{Sean N. Raymond\altaffilmark{1,2}, Thomas R. Quinn\altaffilmark{2}
\&  Jonathan I. Lunine\altaffilmark{3}}

\altaffiltext{1}{Corresponding author: Email -- raymond@astro.washington.edu,\\ phone
 (206) 543-9039, fax (206) 685-0403}
\altaffiltext{2}{Department of Astronomy, University of Washington, Box 351580,
Seattle, WA 98195}
\altaffiltext{3}{Lunar and Planetary Laboratory, The University of Arizona,
Tucson, AZ 85287.}

\begin{abstract}
We present results from 42 simulations of late stage planetary accretion,
focusing on the delivery of volatiles (primarily water) to the terrestrial planets.  Our
simulations include both planetary ``embryos'' (defined as Moon to Mars sized protoplanets)
and planetesimals, assuming that the embryos formed via oligarchic growth.  We
investigate volatile delivery as a function of Jupiter's mass, position and
eccentricity, the position of the snow line, and the density (in solids) of
the solar nebula. 

In all simulations, we form 1-4 terrestrial planets inside 2 AU, which vary in mass and
volatile content.  In 42 simulations we have formed 43 planets between 0.8 and
1.5 AU, including 11 ``habitable'' planets between 0.9 and 1.1 AU.  These
planets range from dry worlds to ``water worlds'' with 100+
oceans of water (1 ocean = 1.5 $\times$ 10$^{24}$ g), and vary in mass
between 0.23$\mearth$ and 3.85 $\mearth$.

There is a good deal of stochastic noise in these simulations, but the most
important parameter is the planetesimal mass we choose, which reflects the
surface density in solids past the snow line.  A high density in this region
results in the formation of a smaller number
of terrestrial planets with larger masses and higher water content, as compared
with planets which form in systems with lower densities.

We find that an eccentric Jupiter produces drier terrestrial planets with
higher eccentricities than a circular one.  In cases with Jupiter at 7 AU, we
form what we call ``super embryos,'' 1-2 $\mearth$ protoplanets which can serve as
the accretion seeds for 2+ $\mearth$ planets with large water contents.

\end{abstract}

\noindent {\bf key words:} 
planetary formation --
extrasolar planets --
origin, solar system --
cosmochemistry

\section{Introduction}

There is a paradox in the definition of the habitable zone with respect to the
presence of liquid water.  Imagine a planet at the right distance from a star to have
stable liquid water on its surface, supported by a modest greenhouse effect.
Nebular models and meteorite data suggest that the local environment during
the formation of this planet was sufficiently hot to prevent hydration of the
planetesimals and protoplanets out of which the planet was formed (Morbidelli
et al., 2000).  That is, the local building blocks of this ``habitable''
planet were devoid of water.  How, then, could this planet acquire water and
become truly habitable?  Delivery of water-laden planetesimals from colder
regions of the disk is one solution, but it implies that the habitability of
extrasolar planets depends on the details of their final assembly, with
implications for the abundance of habitable planets available for Terrestrial
Planet Finder (TPF) to discover.

In the current paradigm of planet formation four dynamically distinct stages
are envisioned (Lissauer, 1993):

\begin{description}

\item{\bf Initial Stage:} Grains condense and grow in the hot
  nebular disk, gradual settling to the mid-plane.  The
  composition of the grains is determined by the local temperature of
  the nebula.
  Gravitational instability among the grains is resisted owing to
  continuous stirring by convective and turbulent motions.

\item{\bf Early Stage:} Growth of grains to km-sized planetesimals occurs via
  pairwise accretion in the turbulent disk.  Planetesimals initially
  have low eccentricities ($e$) and inclinations ($i$) due to gas
  drag.

\item{\bf Middle Stage -- Oligarchic Growth:}
  ``Focused merging'' leads to agglomeration of planetesimals into Moon-to
  Mars-sized ``planetary embryos.''  Possible runaway accretion and subsequent energy
  equipartition (dynamical friction) may lead to polarization of the
  mass distribution: a few large bodies with low $e$ and $i$ in a
  swarm of much smaller planetesimals with high $e$ and $i$.
  The timescale for this process correlates inversely with heliocentric
  distance.  Simulations of oligarchic growth by Kokubo \& Ida (2000)
  suggest that planetary embryos form in $<$ 1 Myr at 1 AU, in $\sim$40 Myr at
  5 AU, and in $>$ 300 Myr past 10 AU.  Since the giant planets are
  constrained to have formed within 10 Myr, embryos could only have formed in
  the innermost solar system within that time.  Thus, we expect that at the
  time of the formation of Jupiter, the inner terrestrial region was dominated by $\sim$
  30-50 planetary embryos while the asteroid belt consisted of a large number of
  $\sim$1 km planetesimals.   

\item{\bf Late Stage:} Once runaway accretion has terminated due to
  lack of slow moving material, planetary embryos and planetesimals gradually
  evolve into crossing orbits as a result of cumulative gravitational
  perturbations.  This leads to radial mixing and giant impacts until
  only a few survivors remain.  The timescale for this process is $\sim 10^8$ yr.

\end{description}

Until recently, a leading hypothesis for the origin of Earth's water was
the ``late veneer'' scenario, in which the Earth
formed primarily from local material, and acquired its water at later times
from a large number of cometary impacts.  The D/H ratio of three comets has
been measured to be 12
times higher than the protosolar value (Balsiger et al., 1995; Meier et
al., 1998; Bockelee-Morvan et al.,1998) , and
roughly twice as high as the terrestrial oceanic (roughly the chondritic) value.  At
most 10\% of the Earth's water came from a cometary source.

Morbidelli et al. (2000) proposed that the bulk of the Earth's water may have
come from the asteroid belt in the form of planetary embryos.  The proto-Earth
accreted several embryos from outside its local region, including a few from
past 2.5 AU, which delivered the bulk of the Earth's water.  In this model the
Earth accreted water since its formation, in the form of an early
bombardment of asteroids and comets, a few large ``wet'' planetary embryos,
and continual impacts of small bodies over long timescales.  This scenario
explains the D/H ratio of Earth's water in the context of late-stage planetary
accretion.  

Morbidelli et al. (2000) assumed that oligarchic growth took place
throughout the inner solar system, with planetary embryos out to 4 AU.  This
neglects Jupiter's strong gravitational influence on the
oligarchic growth process in the asteroid belt, as well as planetesimal-embryo
interactions.  Several other authors (e.g. Chambers 2001; Chambers \& Cassen
2002) subsequently have
numerically formed terrestrial planets, including both planetary embryos
and planetesimals in their
simulations.  Their initial conditions often seem {\it ad
hoc}, and not based on the state of the protoplanetary disk at the end of
oligarchic growth, in particular the radial dependence of the planetary embryo
formation timescale.

Chambers (2003) used a statistical \"Opik-Arnold method to test planet formation
in a number of scenarios, including some which are similar to those we present in
this paper.  The advantage of his statistical method is its low computational
expense relative to N-body simulations, allowing the quick exploration
of a large parameter space.  Its drawback is the
difficulty of implementing realistic dynamics.  Therefore, N-body
simulations like those
we present here may be used to ``calibrate'', and thereby complement
the statistical simulations.

In this paper, we characterize the process of terrestrial planet formation
and volatile delivery as a function of several parameters of the
protoplanetary system.  Our initial conditions attempt to realistically
describe the protoplanetary disk at the beginning of late-stage accretion.  We
do not limit ourselves to our own solar system, and
focus on the formation of planets within the habitable zone of their parent
stars.  The parameters we vary in our simulations are (i) Jupiter's mass,
(ii) eccentricity, (iii) semimajor axis and (iv) time of formation, (v) the
density in solids of the protoplanetary disk and (vi) the location of the snow
line.  

Section 2 describes our initial conditions and numerical methods.  Section 3
presents our results, which are discussed in Section 4, including application
to the NASA's Terrestrial Planet Finder (TPF) mission.  Section 5 concludes
the paper.

\section{Model}

\subsection{Initial Conditions}

The timescale for the formation of planetary embryos from
planetesimals correlates with heliocentric distance (Kokubo \& Ida 2000).  We
wish to accurately describe the state of the protoplanetary disk at the time
when Jupiter formed.  (Note that we use the term ``Jupiter'' to represent the
gas giant planet in each planetary system, the majority of which
differ from our Solar System.  )
We assume that oligarchic growth has taken
place in the inner solar system, from 0.5 AU to the 3:1 mean motion resonance with Jupiter,
located at 2.5 AU in our solar system.  The timescale for embryos to
form out to 2.5 AU is roughly 10 Myr, which is also the upper bound for
Jupiter formation (Briceno et al., 2001).  Past the 3:1 resonance, we expect the gravitational
influence of Jupiter to perturb the oligarchic growth process, through
the clearing out of resonances and other dynamical excitation of planetesimals.

Between the 3:1 resonance and Jupiter we assume that the mass in solids is in
the form of planetesimals.  Since we are computationally limited to a small
number of particles ($\sim$ 200) we can not accurately represent the billions
of planetesimals in this region.  We treat this problem in two ways: (i) All
the mass in the asteroid belt is divided into N ``super-planetesimals'' with
M$_{planetesimal}$ = M$_{ast}$/N, where M$_{ast}$ is the total mass in the region.  (ii)
The planetesimal mass is fixed at 0.01 Earth masses (M$_{\oplus}$).  In
both cases the number of planetesimals is $\sim$150, and they are distributed
as N $\propto$ r$^{-1/2}$, corresponding to the annular mass in a disk with
surface density $\Sigma \propto$ r$^{-3/2}$.  In case (i), the entire mass in
the region (M$_{ast}$) is accounted for, but the ``planetesimals'' have masses
$\sim$ 0.1 M$_{\oplus}$ ($\simeq$ M$_{Mars}$).  In case (ii), the planetesimal
masses are somewhat more realistic, but we do not account for the entire
mass inventory in the region.  Case (i) will provide a rough upper limit on
the mass in volatiles delivered to the terrestrial planets, while case (ii)
corresponds to a lower limit.  Case (ii) only accounts for $\sim$ 10\% of the
mass in the asteroid region, as predicted by Eq.(1).  Therefore, these
simulations are also a test of the effects of the surface density past
the snow line on the forming terrestrial planets and their composition.

We use a two-tiered
surface density profile similar to Kokubo \& Ida (2000), which reflects an
increase in surface density due to the condensation of water immediately
past the snow line:

\begin{equation}
\Sigma(r) = \left\{ \begin{array}{ll}
                   \Sigma_{1} r^{-3/2}, & \mbox{r $<$ snowline} \\
		   \Sigma_{snow} \left( \frac{r}{5 AU}\right) ^{-3/2}, & \mbox{r $>$ snowline}
		   \end{array}
\right. 
\end{equation}

The ``feeding zone'' of a planetary embryo is an annulus with
width comparable to the embryo's Hill Radius,

\begin{equation}
R_{H} = a \left( \frac{M}{3M_{\odot}} \right)^{1/3}
\end{equation}

where a is the embryo's semimajor axis, M is its mass, and M$_{\odot}$ is the
solar mass.  The mass in the feeding zone of an embryo is

\begin{equation}
M = 2\pi a \Sigma R_{H}
\end{equation}


Assuming that the mass of an embryo is proportional to the mass in the
feeding zone, it follows from the surface density profile in Eq.(1)
that the mass of a
planetary embryo M$_{embryo} \propto$ $a^{3/4}$.  Simulations of the formation of
embryos from planetesimals (Kokubo \& Ida 2000) show that they
typically form with separations of 5-10 mutual Hill Radii, rather than immediately
adjacent to each other in heliocentric distance, where the mutual Hill
Radius of
bodies 1 and 2 is defined as 

\begin{equation}
R_{H,m} = \left( \frac{a_{1} + a_{2}}{2} \right) \left( \frac{M_{1} + M_{2}}{3M_{\odot}} \right)^{1/3}
\end{equation}

We therefore space the planetary embryos inside the 3:1
resonance by $\Delta$
mutual Hill radii (i.e. $a_{n+1}$ = $a_{n}$ + $\Delta$R$_{H,m}$),
with $\Delta$ varying randomly between 5 and 10.  So the mass of an
embryo increases as 

\begin{equation}
M_{embryo} \propto a^{3/4} \Delta^{3/2} \Sigma_{1}^{3/2}.
\end{equation}

A snowline at 2.5 AU corresponds, by coincidence,
to the location of the 3:1 resonance for a Jupiter at 5.2 AU.  This implies
that, in our solar system, the local surface density increase at the
snowline implied  by Eq.(1) is not reflected in the masses of planetary
embryos, as they only form interior to the snow line.  If Jupiter's semimajor
axis is less than 5.2 AU the situation is the same, as the 3:1
resonance is located interior to the snowline.  However, if Jupiter's
orbit is larger than 5.2 AU, then the 3:1 resonance is exterior to the
snowline, and there is a region in which oligarchic growth has
taken place in a high density environment.  This results in the formation
of what we call ``super embryos'', which can have masses as large as 2
M$_{\oplus}$.  These
large icy bodies probably did not form in our solar system, but their
dynamical presence can affect terrestrial planet formation interior to their
orbits (see Section 3).

In our simulations, we vary the following parameters:
\begin{enumerate}
\item The semimajor axis of Jupiter's orbit: $a_J$ = 4, 5.2 or 7 AU.
\item Jupiter's Mass: M${_J}$ = 10 M$_{\oplus}$, 1/3 M$_{J,r}$, M$_{J,r}$ or 3 M$_{J,r}$, where
M$_{J,r}$ is Jupiter's real mass of 318 Earth masses.
\item The eccentricity of Jupiter's orbit: $e_J$ = 0, 0.1, or 0.2.
\item The location of the snow line: 2 or 2.5 AU.  A snowline at 2 AU results
in the formation of ``super embryos'' for Jupiter at 5.2 AU.
\item The surface density of solids: $\Sigma_{1}$ = 8-10 g cm$^{-2}$,
$\Sigma_{snow}$ = 3-4 g cm$^{-2}$.  Note that the minimum mass solar nebula has
$\Sigma_{1}$ = 6 g cm$^{-2}$.   
\item The planetesimal mass (exterior to the 3:1 resonance): M$_{planetesimal}$ =
M$_{ast}$/N$_{planetesimal}$ or M$_{planetesimal}$ = 0.01 M$_{\oplus}$, where N$_{planetesimal}$ =
150. 
\item The time of Jupiter formation: Jupiter forms at t=0, or starts as a 10
M$_{\oplus}$ seed whose mass increases to M$_{J}$ at t = 10 Myr.  Note that in
three simulations Jupiter's mass was not increased to M$_{J}$, but left at 10
M$_{\oplus}$.  
\end{enumerate}

\begin{figure}
\centerline{\psfig{figure=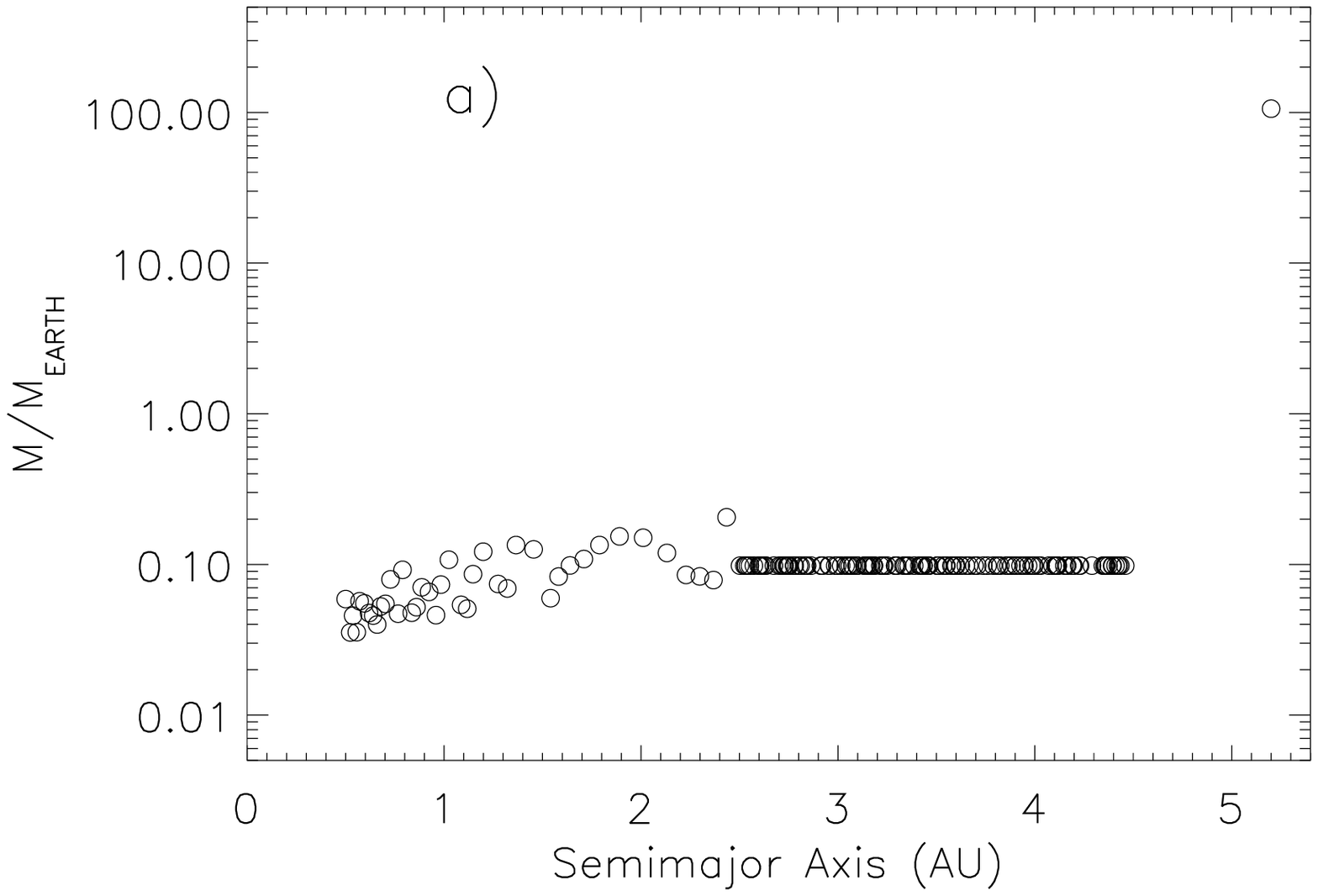,height=8cm}}
\centerline{\psfig{figure=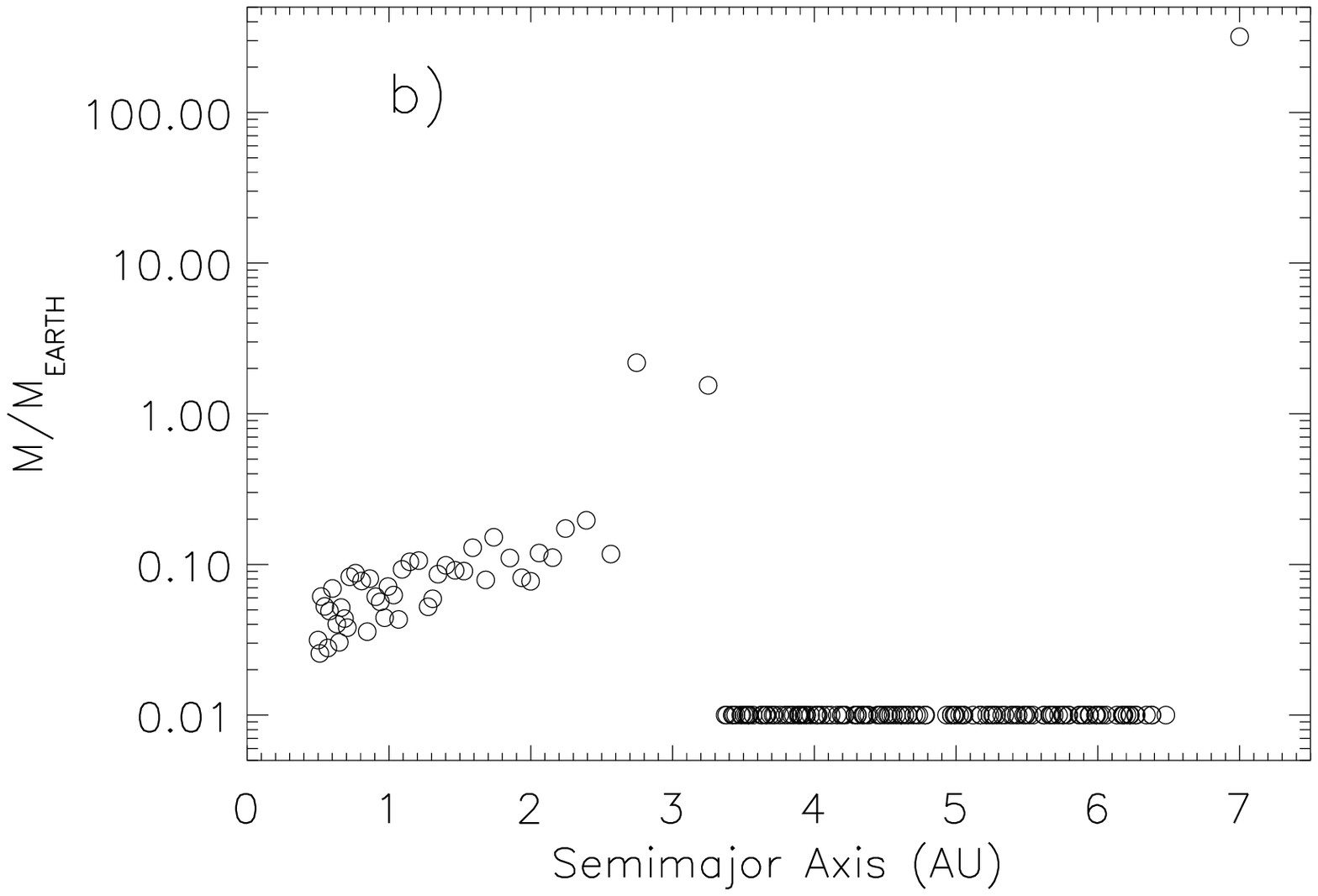,height=8cm}}
\caption{Initial conditions for two simulations.  In the simulation from panel a),
Jupiter's semimajor axis is 5.2 AU and its mass is 1/3 of its real mass.
Planetary embryos are present inside the 3:1 resonance at 2.5 AU, and their
masses are increasing with semimajor axis.  As the separation of embryos
$\Delta$ varies randomly between 5 and 10 mutual Hill radii, the embryos mass
fluctuates according to Eq.(5).  In panel a) the planetesimal mass is $\sim$
0.1 M$_{\oplus}$, and all of the mass in the asteroid belt is contained in
these 150 planetesimals.  In the simulation from panel b), Jupiter's semimajor
axis is 7 AU, and its mass is equal to its current mass.  In this case the 3:1
resonance lies outside the snowline, so there is a region between 2.5 and 3.36
AU in which 1-2 M$_{\oplus}$ ``super embryos'' have formed.  The planetesimal
mass in this simulation is 0.01 M$_{\oplus}$.  Table 1 summarizes the initial
conditions for all 42 simulations.}
\label{fig:init}
\end{figure}

Figure~\ref{fig:init} shows the initial distribution of planetary embryos and
planetesimals for two simulations, showing the range in variation of these
parameters.  Table 1 lists the initial conditions for all 42 simulations.

\scriptsize
\begin{deluxetable}{ccccccccc}
\tablewidth{0pt}
\tablecaption{Initial Conditions for 42 Simulations}
\renewcommand{\arraystretch}{.6}
\tablehead{
\colhead{Simulation } &  
\colhead{$a_{J} (AU)$} & 
\colhead{$e_{J}$} &
\colhead{$M_{J}$ (M$_{J,r}$)\tablenotemark{b}} &  
\colhead{$t_{J}$ (Myr)\tablenotemark{b}} &
\colhead{$M_{pl}$ ($\mearth$)\tablenotemark{c}} & 
\colhead{$\Sigma_1$ (g cm$^{-2}$)}  & 
\colhead{$\Sigma_{snow}$ (g cm$^{-2}$)} & 
\colhead{snow line (AU)} }
\startdata

   1   & 4.0   & 0.0   &  1   &   0   &  0.09   & 10   &  4   & 2.5  \nl
   2   & 4.0   & 0.0   &  1   &   0   &  0.09   & 10   &  4   & 2.5  \nl
   3   & 4.0   & 0.0   &  1   &   0   &  0.09   & 10   &  4   & 2.5  \nl
   4   & 5.2   & 0.0   &  1   &   0   &  0.09   & 10   &  4   & 2.5  \nl
   5   & 5.2   & 0.0   &  1   &   0   &  0.09   & 10   &  4   & 2.5  \nl
   6   & 5.2   & 0.0   &  1   &   0   &  0.09   & 10   &  4   & 2.5  \nl
   7   & 7.0   & 0.0   &  1   &   0   &  0.11   & 10   &  4   & 2.5  \nl
   8   & 7.0   & 0.0   &  1   &   0   &  0.11   & 10   &  4   & 2.5  \nl
   9   & 7.0   & 0.0   &  1   &   0   &  0.11   & 10   &  4   & 2.5  \nl
  10   & 5.2   & 0.0   &  1   &   0   &  0.01   & 10   &  3   & 2.5  \nl
  11   & 5.2   & 0.0   &  1   &   0   &  0.01   & 10   &  3   & 2.5  \nl
  12   & 5.2   & 0.0   &  1   &   0   &  0.01   &  8   &  3   & 2.5  \nl
  13   & 5.2   & 0.0   &  1   &   0   &  0.01   &  8   &  3   & 2.5  \nl
  14   & 7.0   & 0.0   &  1   &   0   &  0.01   &  8   &  3   & 2.5  \nl
  15   & 7.0   & 0.0   &  1   &   0   &  0.01   &  8   &  3   & 2.5  \nl
  16   & 4.0   & 0.0   &  1   &   0   &  0.01   &  8   &  3   & 2.5  \nl
  17   & 4.0   & 0.0   &  1   &   0   &  0.01   &  8   &  3   & 2.5  \nl
  18   & 5.2   & 0.0   &  1   &   0   &  0.01   &  8   &  3   & 2.0  \nl
  19   & 5.2   & 0.0   &  1   &   0   &  0.01   &  8   &  3   & 2.0  \nl
  20   & 7.0   & 0.0   &  1   &   0   &  0.01   &  8   &  3   & 2.0  \nl
  21   & 7.0   & 0.0   &  1   &   0   &  0.01   &  8   &  3   & 2.0  \nl
  22   & 5.2   & 0.0   &  1/3   &   0   &  0.01   & 10   &  4   & 2.5  \nl
  23   & 5.2   & 0.0   &  1/3   &   0   &  0.01   & 10   &  4   & 2.5  \nl
  24   & 5.2   & 0.0   &  1/3   &   0   &  0.10   & 10   &  4   & 2.5  \nl
  25   & 5.2   & 0.0   &  1/3   &   0   &  0.10   & 10   &  4   & 2.5  \nl
  26   & 5.2   & 0.0   &  3   &   0   &  0.01   & 10   &  4   & 2.5  \nl
  27   & 5.2   & 0.0   &  3   &   0   &  0.01   & 10   &  4   & 2.5  \nl
  28   & 5.2   & 0.0   &  3   &   0   &  0.10   & 10   &  4   & 2.5  \nl
  29   & 5.2   & 0.0   &  3   &   0   &  0.10   & 10   &  4   & 2.5  \nl
  30   & 5.2   & 0.1   &  1   &   0   &  0.01   & 10   &  4   & 2.5  \nl
  31   & 5.2   & 0.1   &  1   &   0   &  0.01   & 10   &  4   & 2.5  \nl
  32   & 5.2   & 0.1   &  1   &   0   &  0.01   & 10   &  4   & 2.5  \nl
  33   & 5.2   & 0.2   &  1   &   0   &  0.01   & 10   &  4   & 2.5  \nl
  34   & 5.2   & 0.2   &  1   &   0   &  0.01   & 10   &  4   & 2.5  \nl
  35   & 5.2   & 0.2   &  1   &   0   &  0.01   & 10   &  4   & 2.5  \nl
  36   & 5.5   & 0.0   &  1   &  10   &  0.01   & 10   &  4   & 2.5  \nl
  37   & 5.5   & 0.0   &  1   &  10   &  0.01   & 10   &  4   & 2.5  \nl
  38   & 5.5   & 0.0   &  1   &  10   &  0.01   & 10   &  4   & 2.5  \nl
  39   & 5.5   & 0.0   &  1   &  10   &  0.01   & 10   &  4   & 2.5  \nl
  40   & 5.5   & 0.0   &  0.03   &   0   &  0.01   & 10   &  4   & 2.5  \nl
  41   & 5.5   & 0.0   &  0.03   &   0   &  0.01   & 10   &  4   & 2.5  \nl
  42   & 5.5   & 0.0   &  0.03   &   0   &  0.01   & 10   &  4   & 2.5  \nl

\enddata

\tablenotetext{a}{Jupiter's mass, in units of its real mass M$_{J,r}$.  In
simulations 40-42, Jupiter's mass is 10 $\mearth$.}
\tablenotetext{b}{Time of Jupiter formation.  For simulations with $t_{J}$ =
10 Myr, Jupiter began the simulation as a 10 $\mearth$ accretion seed and was
inflated to Jupiter's real mass at 10 Myr.}
\tablenotetext{c}{The mass of a planetesimal, in earth masses.  Referred to as
m$_{planetesimal}$ in text.}
\end{deluxetable}

\clearpage
\normalsize

\subsection{Water Content}

The water content of planetesimals in a given planetary system depends in a
complex way upon a range of factors including the mass and evolutionary
characteristics of the protoplanetary disk, overall metallicity of the
molecular cloud clump from which the star is forming, and the positions,
masses and timings of formation of the system's giant planets. Although in
this paper we track the dynamical history of planetesimals for a given set of
giant planet orbital parameters, our model is not capable of determining the
water content of the planetesimals versus semi-major axis, in part because
this is determined long before the stage at which planetesimals grow to the
size of the Moon or larger. However, it should be said that even nebular
models concerned with earlier stages of planetesimal growth lack the fidelity
to create {\it ab initio} a reliable map of water content versus semi-major axis. In
this paper we assume a distribution of water content versus semi-major axis
for the planetesimals, in all of our model systems, based on the meteorite
data for our own solar system. The stochastic delivery of water to the
terrestrial planets that is one of the main conclusions of this paper argues
against any more detailed attempt to scale the water abundance for any given
system according to disk parameters, for the resulting delivery of water will
remain stochastic.  

\begin{figure}[p]
\centerline{\psfig{figure=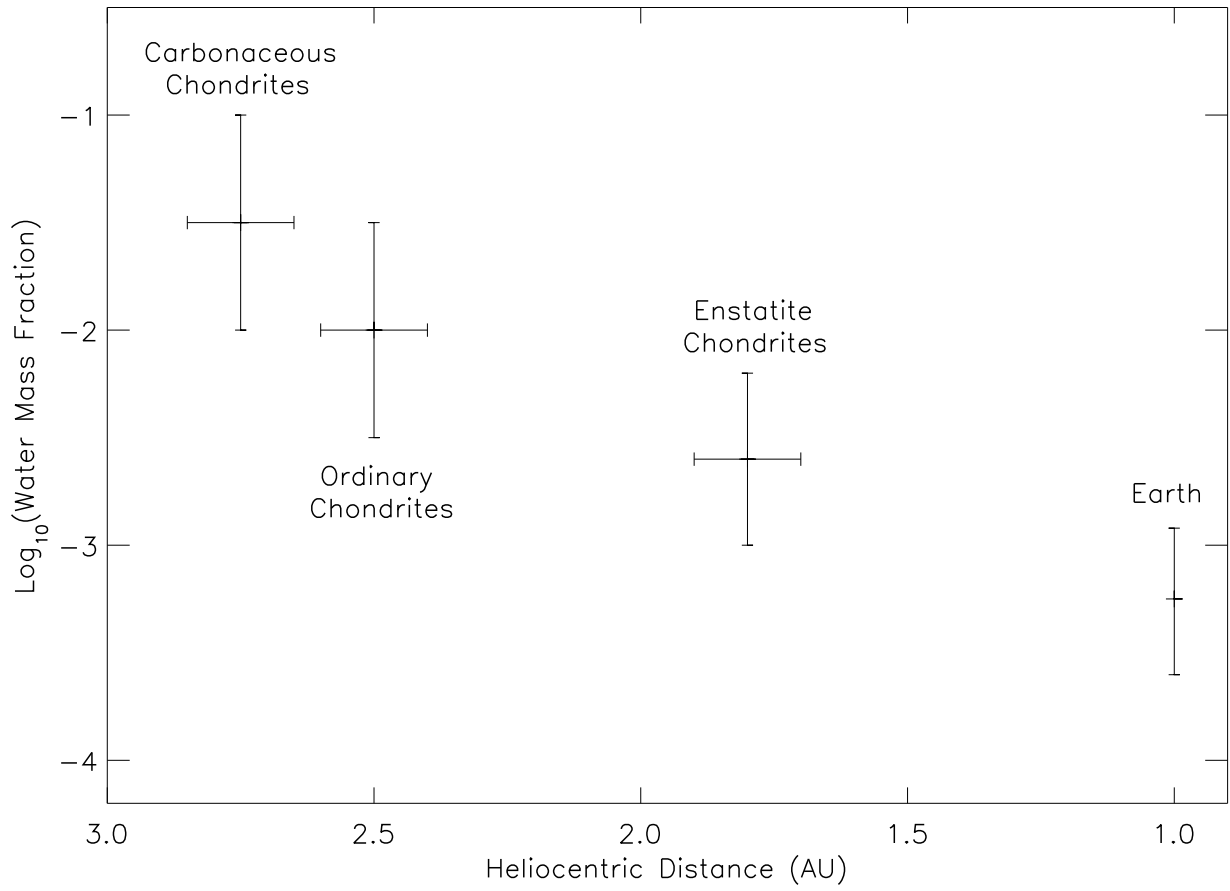}}
\caption{Water content of various types of chondrites in our solar system,
with approximate values for the positions of the parent bodies. Water
percentages from Abe et al. (2000).} 
\label{fig:lun1}
\end{figure}

Figure~\ref{fig:lun1} gives the range of water content, by mass, of three types of
chondritic meteorites compared to an estimate for the Earth. Carbonaceous
chondrites are the most water rich of meteorites, with water content up to
nearly 10\% by mass. The ordinary chondrites are significantly lower, roughly
by an order of magnitude, and the enstatite chondrites are somewhat lower
still  (Abe et al. 2000). The locations within our solar system from which the
parent bodies of the meteorites were derived is highly uncertain, though it is
generally thought that the carbonaceous chondrites came from the region beyond
2.5 AU, while the ordinary and enstatite chondrites were formed further
inward. Earth, for which no primitive counterpart exists in the meteorite
record discovered to date (Drake and Righter 2002), is quite dry with a water
content around 0.03-0.1\% by mass (Abe et al. 2000). Much of this may have
been added from planetesimals at larger semi-major axes than 1 AU (Morbidelli
et al. 2000), consistent with the dynamical results we present below. Also,
the lunar-forming impactor appears to have been extremely dry, with water
content much less than that of the Earth (Abe et al. 2000). Hence, planetesimals at
1 AU could have been orders of magnitude dryer than the water content of the
Earth today. 

Hydration of the silicates in meteorites is thought to have occurred inside a
parent body rather than in the solar nebula, due to the very long hydration
timescales in the solar nebula (Fegley, Jr., 2000).  This implies that
hydrated bodies could only form by accreting ice, which then
reacted as liquid water with the anhydrous minerals when sufficiently heated (by
radioactive, frictional or collisional heating).  The current distribution of
water-rich vs dry classes of asteroids may be a
record of the position of the snow line in the solar nebula (neglecting
possible orbital migration of asteroids).  As mentioned above, this division
occurs at roughly 2-2.5 AU in our solar system, and corresponds to a nebular
temperature of $\sim$ 180 K.  A density increase immediately past the snow
line is expected due to the ``cold trap'' effect (Stevenson \& Lunine, 1988),
and is expressed in equation (1).  



As a baseline, we divide each of the planetary
systems into three regions according to water
content-planetesimals beyond 2.5AU have 5\% water by mass, those inward of 2
AU have water content of 0.001\% by mass, and between 2-2.5 AU lie
planetesimals with intermediate water content of 0.1\% by mass. This
distribution of water among the planetesimals can be seen in the first panel
of the accretion simulations we show below (Fig.~\ref{fig:aet1}, for example). The
results of our calculations are such that the intermediate planetesimal class,
in terms of water content, does not affect the overall conclusions regarding
delivery of water to Earth. Thus, one can think of the simulations as positing
two regions---one water-rich beyond 2-2.5 AU and one water poor inward of
that---and then following through the collisional history of the planetesimals
the delivery of water to the final few terrestrial planets remaining at the
end of each simulation.  

\subsection{Numerical Method}

We integrate all simulations for 200 Myr using Mercury (Chambers 1999).  We use
the hybrid integrator, which uses a second-order mixed variable
symplectic algorithm when objects are separated by more than 3 Hill radii, and
a Burlisch-Stoer method for closer encounters.  We use a 6 day timestep, in
order to have 15 timesteps per orbit for the innermost orbits in our
initial conditions at 0.5 AU. Our simulations conserve energy to better than 1 part in
10$^{4}$, and angular momentum to 1 part in 10$^{11}$.  Collisions conserve
linear momentum and do not take fragmentation into account.  These simulations were
run on desktop PCs, each taking roughly 1 month of CPU time on a 700 MHz
machine.  The three simulations with M$_{J}$ = 10 M$_{\oplus}$ each required
4-6 months of CPU time.

\section{Results}

In this section we first discuss the details of one simulation,
describing the physical processes which apply to all of our simulations.  We
then show the dependences of the terrestrial planets we form on planetary
system parameters, and statistically examine the terrestrial planets which
have formed.  We then address the issues of water content and habitability.

\subsection{One Simulation}

\begin{figure}
\psfig{figure=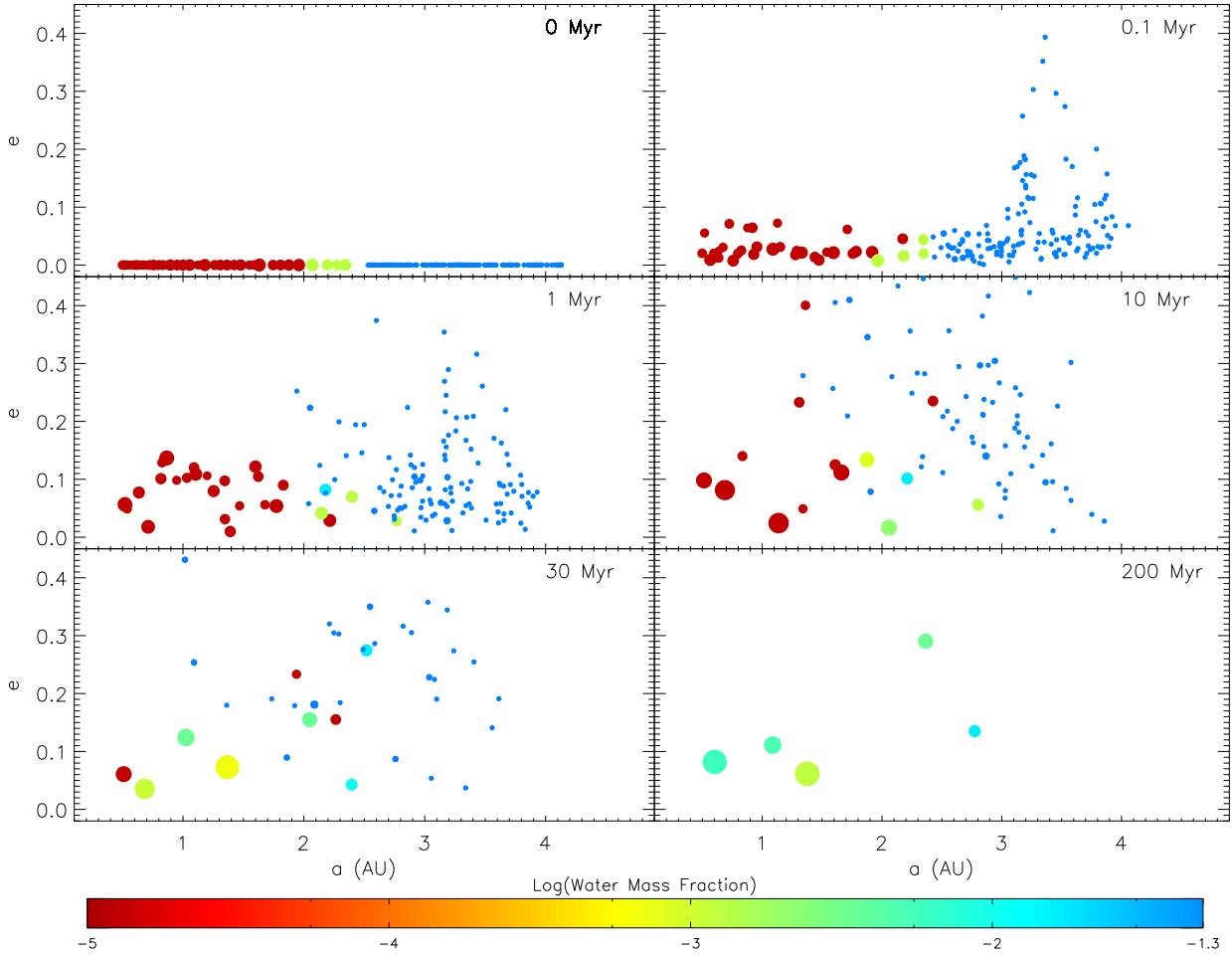,height=13cm,angle=90}
\caption{Snapshots in the evolution of a simulation with Jupiter at 5.2 AU
with zero eccentricity, and a planetesimal mass of 0.01 M$_{\oplus}$
(simulation 10: see table 1 for details).  The
size of each object is proportional to its mass$^{(1/3)}$ (but does not
represent the actual physical size), and the
color of each object corresponds to its water mass fraction.  Note that the
wettest objects have water mass fractions of log$_{10}$(5\%) = -1.3.  See text
for discussion.}
\label{fig:aet1}
\end{figure}

Figure~\ref{fig:aet1} shows six snapshots in time of the evolution of one
simulation (simulation 10; see Table 1) with $a_{J}$ = 5.2 AU, $e_J$ = 0, M$_{J}$ = M$_{J,r}$, and
0.01 M$_{\oplus}$ planetesimals.  As mentioned in Section 2.2, we assume that objects are dry
(0.001\% water) if they form interior to 2 AU, wet (5\% water) if they form
exterior to 2.5 AU, and moderately wet (0.1\% water) in between.  These
compositional constraints are reflected in the initial conditions of
Fig.~\ref{fig:aet1}.  

Jupiter begins to excite the eccentricities of planetesimals in the
asteroid belt, especially those which are located close to mean motion
resonances.  The 0.1 Myr snapshot in Fig.~\ref{fig:aet1} clearly shows the
2:1 resonance at 3.28 AU, the 3:2 resonance at 3.97 AU, and a hint of the 5:3
resonance at 3.7 AU.  Meanwhile, the planetary embryos' eccentricities are
being excited to a smaller degree by their mutual gravitational pulls.  As the
eccentricities of planetesimals increases, their orbits become crossing, and
the probability of a collision or a close encounter with Jupiter increases.
A collision between two objects with different semimajor axes typically occurs
when one is at perihelion and the other is at aphelion.  The alignment of
their velocities implies that a collision tends to circularize the orbit of the
remaining agglomeration.  Therefore, there are two likely end states for a
planetesimal in resonance with Jupiter: (i) The planetesimal will collide with
an object with a smaller semimajor axis, effectively moving the planetesimal
closer to the Sun, outside of the resonance with Jupiter.  (ii) The planetesimal's
eccentricity increases to the point where it will have a close encounter
with Jupiter and likely be ejected from the system.

By t = 1 Myr in Fig.~\ref{fig:aet1} radial mixing has begun at the
boundary between the dry planetary embryos and the wet planetesimals.  Over
the next 10 Myr planetary embryos accrete to form larger protoplanets, as
planetesimals slowly diffuse inside 2 AU.  Between 10 and 30 Myr, most
protoplanets have accreted some material from past the snow line. At the end
of the simulation (t = 200 Myr), three terrestrial planets have been formed
inside 2 AU, and two smaller bodies reside in the asteroid belt.  No
planetesimals remain in the asteroid belt.  Each surviving body contains a
mixture of dry and wet material, to varying degrees.  The eccentricities of the
planets is $\sim$ 0.1 inside 2 AU, and larger for the asteroidal planets.  Two
roughly Earth mass planets have formed, one inside and one outside 1 AU.  One
planet has formed in the habitable zone, at 1.08 AU, but is significantly less
massive at 0.4 M$_{\oplus}$, and has a relatively high eccentricity of 0.1,
whereas the more massive planets have slightly smaller eccentricities.  The
exact parameters of the surviving planets for all simulations are listed in Table 1. 

\begin{figure}
\centerline{\psfig{figure=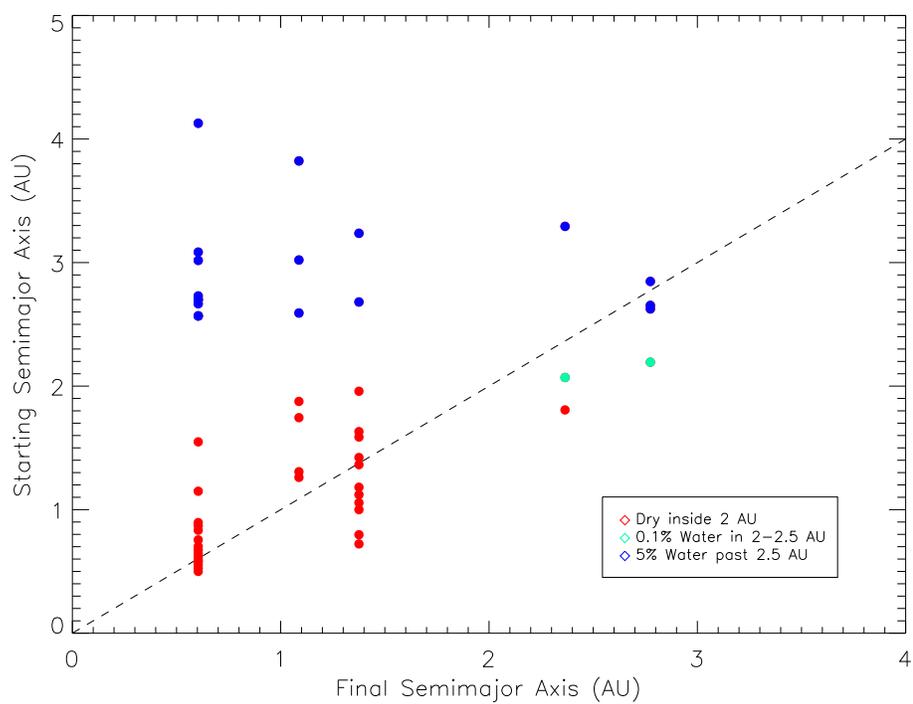,height=10cm,angle=90}}
\caption{Starting vs final semimajor axes for all objects which incorporated
into the five surviving bodies from Fig.~\ref{fig:aet1}.  The dashed line
is where the starting and final values are equal.  See text for discussion.}
\label{fig:aa1}
\end{figure}

For each of the five surviving planets from simulation 10, Fig.~\ref{fig:aa1}
shows the origin of every planetary embryo and planetesimal which was accreted
by that planet.  The dashed line indicates where the starting and final
semimajor axes are equal.  An object close to the dashed line had little
radial displacement through the formation process, whereas an object far from
the dashed line exhibited a large radial excursion (e.g. the planetesimals
from past 3 AU which were accreted by the planet at 0.6 AU).  Each of the
planets inside 2 AU was formed from objects throughout the inner solar system,
including at least two planetesimals from the outer asteroid belt, although
the majority of their mass came from local planetary embryos.  Interestingly,
the planet at 1.08 AU formed entirely from material exterior to its final
position.  The two planets in the asteroid belt exhibit a narrower zone of
accretion, but are still made up of material from different regions.

\begin{figure}
\centerline{\psfig{figure=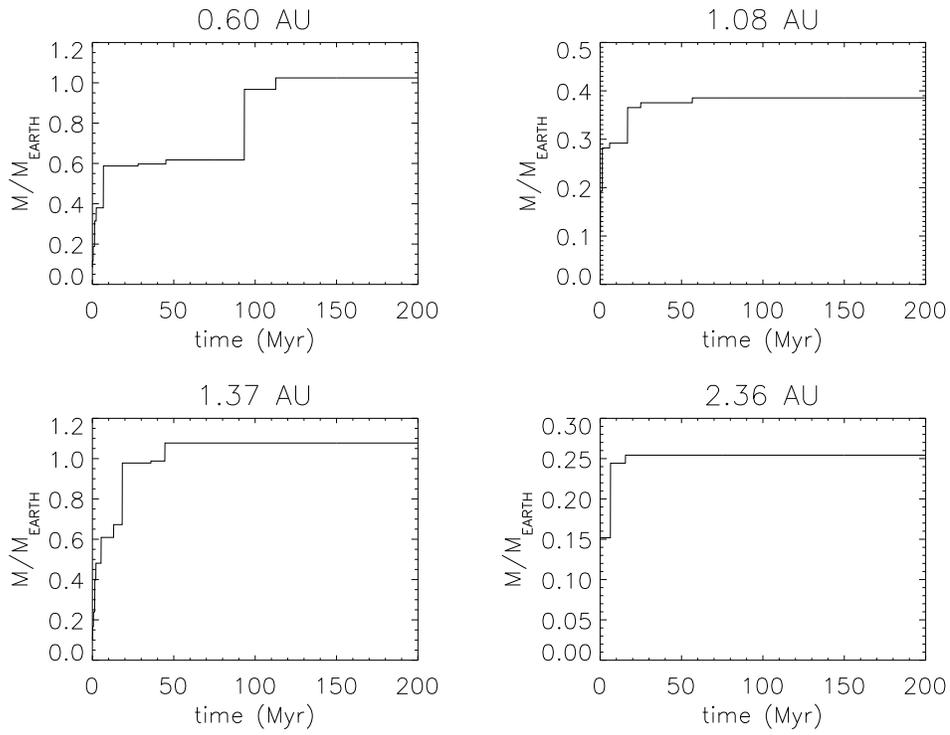,height=10cm,angle=90}}
\caption{Mass vs time for objects in the simulation from
Fig.~\ref{fig:aet1}, labeled by their final semimajor axes.  The three
planets inside 2 AU are included, as well as the innermost object in the
asteroid belt.}
\label{fig:mass1}
\end{figure}

Figure~\ref{fig:mass1} shows the masses of the planets from simulation 10 as a
function of time, labeled by their final semimajor axes.  The three planets
inside 2 AU are included, as well as the
inner of the two asteroid belt planets.  The planets reach half of
their final masses within the first 10-20 Myr, although significant accretion events
occur as late as 100 Myr.  There is a constraint for our own Solar System from
measured Hf-W ratios that both the Moon and the Earth's core were
formed by t $\simeq$ 30 Myr (Kleine et al. 2002; Yin et al. 2002).  This implies
that the Earth's mass was within a factor of 2 of its current mass at that
time.  Simulation 10 satisfies this constraint.  This will be discussed more
generally in Section 4.

\subsection{Dependences on System Parameters}

In all
cases we form 1-4 terrestrial planets with masses from 0.23 to 3.85 
M$_{\oplus}$, which vary in orbital parameters and volatile content.  We
define a terrestrial planet as a surviving body residing inside 2 AU, with a
mass greater than 0.2 $\mearth$.
There is a large variation among the planetary systems which are formed, as
well as significant stochastic variation between simulations with identical
parameters.  We now summarize how the terrestrial planets we form are affected
by planetary system parameters:

{\bf Planetesimal mass:}  The mass of a planetesimal in the outer asteroid
region causes large 
variations in the terrestrial planets which are formed.  Recall that in
Section 2.1 we defined the two planetesimal masses we use in our simulations.
In case (i) the total mass
in the asteroid region is accounted for and the planetesimal mass is $\sim$
0.1 $\mearth$, and in case (ii) the planetesimal mass is fixed at 0.01
$\mearth$.  Cases (i) and (ii) can be thought of as realistic upper and lower
limits to the amount of material in solids past the snow line.

The mean water mass fraction of the planets formed in all case (i) simulations
is 1.7$\times$10$^{-2}$ vs 4$\times$10$^{-3}$ for all case (ii) simulations.
The average planet mass for case (i) is 1.8 $\mearth$ vs 0.9 $\mearth$ for
case (ii).  The mean number of planets per simulation for case (i) is 2.5 vs
3.2 for case (ii), and the total mass in terrestrial planets is 4 $\mearth$
for case (i) vs 2.7 $\mearth$ for case (ii).  The mean eccentricities of all
terrestrial planets for case (i) is 0.14, vs 0.10 for case (ii).

{\bf Surface density:} The effects of surface density should be similar to
those of the planetesimal mass.  However, since we have only covered a
relatively small range in values of $\Sigma_1$ and $\Sigma_{snow}$ in these
simulations, the direct effects are negligible compared with the effects of
the planetesimal mass and the stochastic ``noise'' inherent to the
simulations.  

{\bf Jupiter's mass:} A Jovian planet of larger mass forms a smaller number of
terrestrial planets than a lower-mass body.  The masses of the terrestrial
planets increase slightly with the mass of the Jovian planet, but this is a small effect
compared with the effects of the planetesimal mass.  In our 30 case (ii)
simulations, the total mass incorporated in
terrestrial planets is roughly constant with M${_J}$ at $\sim$ 2.5 $\mearth$, 25\% higher
than in our solar system.  The water content of the terrestrial planets does
not vary significantly with M${_J}$.  

The number of surviving bodies at the
end of the 200 Myr integration (which includes planets and remnant planetesimals and
planetary embryos) increases sharply at small Jupiter masses, with an average
of 54 for a 10 $\mearth$ Jupiter.  Since the algorithm used in Mercury
scales with the number of bodies, $n$, as $n^2$, this explains the large increase in
computational expense for the 10 $\mearth$ Jupiter simulations compared with
those with a more massive gas giant.

\begin{figure}
\centerline{\psfig{figure=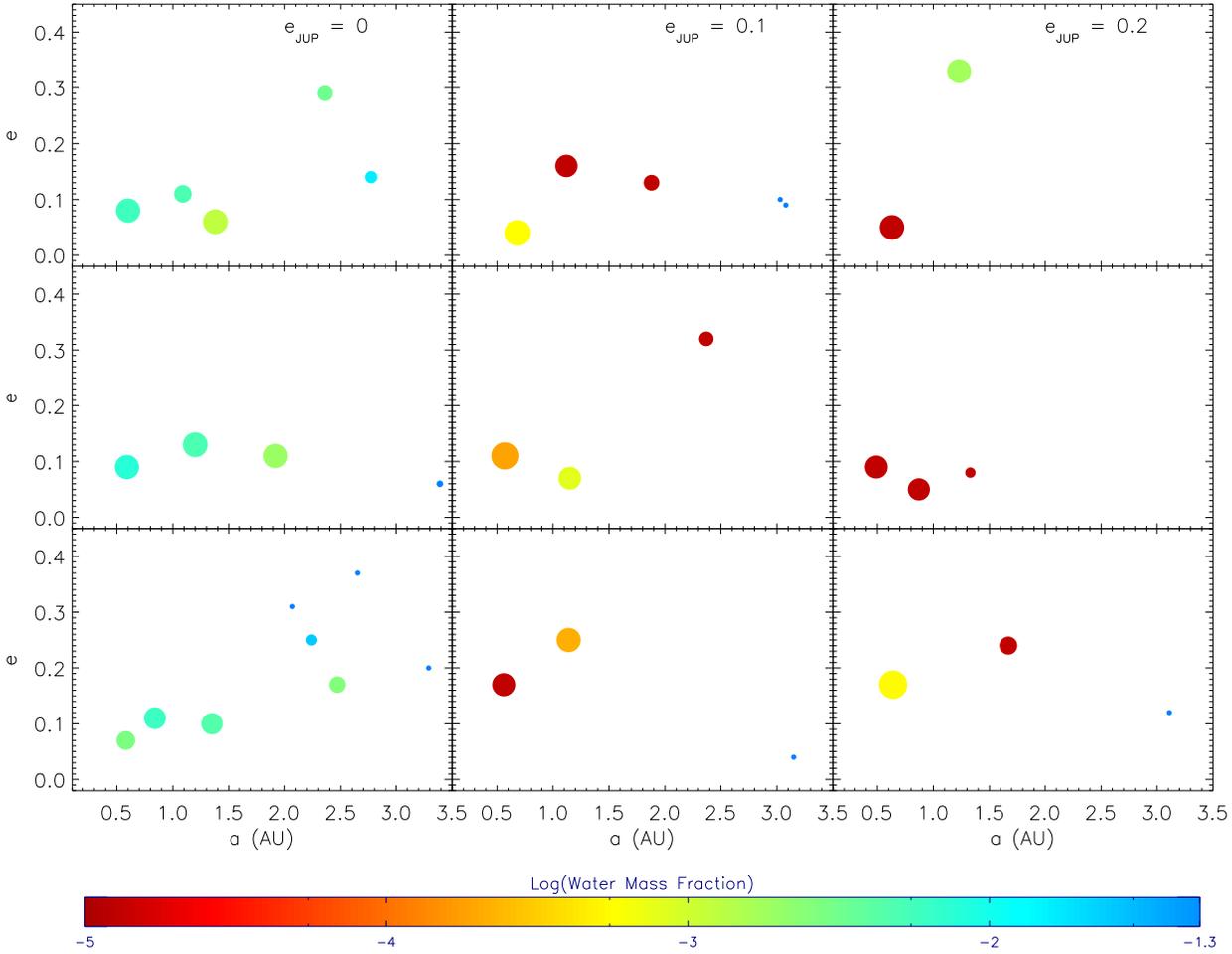,height=13cm,angle=90}}
\caption{The final configuration of nine planetary systems with identical
initial conditions ($a_J$ = 5.2 AU, M$_J$ = M$_{J,r}$, M$_{planetesimal}$ = 0.01
$\mearth$) apart from Jupiter's initial eccentricity, which is the 
same for all simulations in a given column.  Note the dramatic decline in
volatile content for $e_J$ greater than zero.} 
\label{fig:ecc}
\end{figure}

{\bf Jupiter's eccentricity:}  Figure~\ref{fig:ecc} shows the configuration of
nine planetary systems at the end of a 200 Myr integration, including three
with $e_J$ = 0.1 and three with $e_J$ = 0.2.  An eccentric Jupiter clears out
the asteroid region much more quickly than a low eccentricity Jupiter, and forms
volatile-poor terrestrial planets.  The mean water mass fraction of fourteen
terrestrial planets which formed in the six simulations with $e_J$ = 0.1 or
0.2 is 2$\times$10$^{-4}$, vs 8$\times$10$^{-3}$ for all simulations with
$e_J$ = 0.  The mean total mass in terrestrial planets in these six
simulations was 2.0 $\mearth$, as compared with 2.5 $\mearth$ for all case
(ii) simulations with $e_J$ = 0.  In five out of these six simulations the
most massive terrestrial planet
which formed was the innermost.  In addition, the terrestrial planets in these
simulations have higher average eccentricities than those with $e_J$ = 0 (0.14 vs
0.10).

To first order, the effective width of a Jupiter mean motion
resonance scales with Jupiter's eccentricity as $\delta a/a \propto e_J^{1/2}$
(Murray \& Dermott 2001, eqn 8.76).  This is a simplification, as the
phase space structure of resonances is quite complex.  However, this explains
the much more rapid depletion of the asteroid belt in the
case of an eccentric Jupiter, resulting in volatile-poor terrestrial planets.

{\bf Jupiter's semimajor axis:}  We see a weak correlation between higher water
content in terrestrial planets and Jupiter's semimajor axis.  In the case
of a Jupiter at 4 AU, there is simply a smaller reservoir of volatile-rich
material between the snow line and Jupiter as compared with a system with
$a_J$ = 5.2 AU, resulting in drier planets.  In the case of a Jupiter at 7 AU,
this reservoir is much larger.  However, as Jupiter's zone of influence is at
larger heliocentric distances, there is a larger change in energy required to
deliver volatile-rich planetesimals to the terrestrial region.  In addition,
the super embryos which form between the snowline and the 3:1 Jupiter
resonance present a dynamical obstacle for inward-bound planetesimals.
As a result, the innermost terrestrial planet is completely dry in
four out of seven simulations with $a_J$ = 7 AU, in which the mean water mass
fraction of all planets formed inside 1 AU is 1.7$\times$10$^{-3}$, and
3.8$\times$10$^{-2}$ for planets which formed between 1 and 2 AU.  This is not
unexpected, as super embryos immediately past the snowline at 2.5 AU can act
as accretion seeds and occasionally migrate inside 2 AU, resulting in massive
(1.5 to 3+ $\mearth$) planets in the proximity of Mars' current orbit.

{\bf The position of the snow line:}  The most important consequence of a
snowline at 2 AU rather than 2.5 AU is that it lies inside the 3:1 resonance
with a Jupiter at 5.2 AU, resulting in the formation of super embryos in the
annulus between 2 and 2.5 AU.  As mentioned above, these can act as an
obstacle to inward-diffusing planetesimals.  However, in four simulations with
the snowline at 2 AU we see little difference in the terrestrial planets.
We intend to explore this region of parameter space in more detail in future work.

{\bf Time of Jupiter formation:}  We see no significant
physical differences between the terrestrial planets in simulations with a 10
$\mearth$ Jupiter seed which grew to full size at 10 Myr and those which began
with a full size Jupiter.  More simulations are needed to overcome stochastic
noise and small number statistics, in order to establish correlations.

\subsection{Water Content}
Table 2 summarizes the results of the water added to terrestrial planets
within a semi-major axis range of 0.8-1.5 AU from the parent star. This is
larger than the so-called "continuously habitable zone" (CHZ) defined as the
semi-major axis realm for which a planet's mean surface temperature will be
above the water melting point, but not so high as to allow loss of water by
evaporation and subsequent photolysis (Kasting et al. 1993; Kasting 1988). The
CHZ could be as narrow as 0.05-0.1 AU centered on the Earth's orbit, but this
could be too narrow or too wide given our lack of knowledge regarding the
range of habitable planetary environments and the robustness of processes that
buffer stable liquid water environments on planets. A better constraint would
be the search space planned for the Terrestrial Planet Finder (NASA) and
Darwin (ESA) programs designed to detect and characterize Earths around other
stars (see part 4). This search space, for a solar-mass star such as that
which we are considering here, ranges from 1.5 AU (the orbit of Mars in our
solar system) to as close as 0.7 AU (Venus' orbit) for a subset of the stars on
the search list.

\scriptsize
\begin{deluxetable}{ccccccc}
\tablewidth{0pt}
\tablecaption{``Habitable'' Planets with 0.8 AU $<$ a $<$ 1.5 AU {\bf (0.9 AU $<$ a $<$ 1.1 AU)}\tablenotemark{a}}
\renewcommand{\arraystretch}{.6}
\tablehead{
\colhead{Simulation} &  
\colhead{a (AU)} &
\colhead{e} &
\colhead{i (deg)} &
\colhead{Mass ($\mearth$)} & 
\colhead{N(oceans)\tablenotemark{b}} &
\colhead{W.M.F.\tablenotemark{c}}}
\startdata

\bf  1   & \bf  1.09   & \bf  0.11   & \bf   4.1   & \bf  2.02   & \bf   53  & {\bf 6.7$\times$10$^{\bf -3}$}  \nl
\bf  2   & \bf   0.98   & \bf  0.15   & \bf  19.3   & \bf  0.86   & \bf    0
& \bf 1.0$\times$10$^{\bf -5}$  \nl
  3   &  1.24   &  0.10   &   2.7   &  2.03   &   35   & 4.4$\times$10$^{-3}$  \nl
  4   &  1.15   &  0.08   &   7.7   &  1.28   &   36   & 7.1$\times$10$^{-3}$  \nl
  5   &  1.40   &  0.25   &   9.5   &  0.96   &   91   & 2.4$\times$10$^{-2}$  \nl
\bf 10a   & \bf  1.09   & \bf  0.11   & \bf   8.1   & \bf  0.38   & \bf    5
& \bf 3.9$\times$10$^{\bf -3}$  \nl
 10b   &  1.38   &  0.06   &   1.4   &  1.08   &    3   & 9.3$\times$10$^{-4}$  \nl
 11   &  1.20   &  0.13   &   6.5   &  1.09   &   16   & 3.8$\times$10$^{-3 }$ \nl
 12a   &  0.84   &  0.11   &   6.3   &  0.74   &   14   & 4.8$\times$10$^{-3}$  \nl
 12b   &  1.35   &  0.10   &   5.0   &  0.70   &    9   & 3.6$\times$10$^{-3}$  \nl
\bf 13   & \bf  1.00   & \bf  0.02   & \bf   4.5   & \bf  1.27   & \bf   13
& \bf 2.6$\times$10$^{\bf -3}$  \nl
 14   &  1.43   &  0.07   &   1.5   &  1.94   &  297   & 3.8$\times$10$^{-2}$  \nl
 15   &  1.43   &  0.06   &  14.9   &  1.17   &  183   & 3.9$\times$10$^{-2}$  \nl
 16   &  1.43   &  0.04   &   8.3   &  1.14   &    4   & 9.3$\times$10$^{-4 }$ \nl
\bf 17a   & \bf  1.01   & \bf  0.03   & \bf   1.8   & \bf  0.61   & \bf    6
& \bf 2.5$\times$10$^{\bf -3}$  \nl
 17b   &  1.43   &  0.00   &   3.5   &  0.62   &    4   & 1.8$\times$10$^{-3}$  \nl
\bf 19   & \bf  1.00   & \bf  0.17   & \bf   6.8   & \bf  0.71   & \bf    8
& \bf 2.9$\times$10$^{\bf -3 }$ \nl
 19   &  1.44   &  0.10   &   4.3   &  1.02   &   17   & 4.4$\times$10$^{-3 }$ \nl
 20   &  1.48   &  0.10   &   1.6   &  2.82   &  322   & 2.9$\times$10$^{-2}$  \nl
 21   &  1.26   &  0.10   &  14.3   &  1.08   &  154   & 3.6$\times$10$^{-2 }$ \nl
 \bf 22   & \bf  0.92   & \bf  0.03   & \bf   3.7   & \bf  1.25   & \bf   27
 & \bf 5.6$\times$10$^{\bf -3}$  \nl
 23a   &  0.84   &  0.05   &   0.7   &  0.96   &   11   & 3.1$\times$10$^{-3}$  \nl
 23b   &  1.36   &  0.10   &   1.4   &  0.86   &   11   & 3.5$\times$10$^{-3}$  \nl
 24   &  1.32   &  0.16   &   7.8   &  1.77   &  215   & 3.1$\times$10$^{-2}$  \nl
\bf 25   & \bf  1.05   & \bf  0.07   & \bf   5.8   & \bf  3.11   & \bf  292
& \bf 2.4$\times$10$^{\bf -2}$  \nl
 26   &  1.33   &  0.07   &   5.1   &  1.54   &   13   & 2.1$\times$10$^{-3}$  \nl
 27   &  1.21   &  0.06   &   5.9   &  1.07   &   26   & 6.2$\times$10$^{-3}$  \nl
\bf 28   & \bf  0.96   & \bf  0.14   & \bf   5.1   & \bf  3.85   & \bf  352
& \bf 2.3$\times$10$^{\bf -2 }$ \nl
 29   &  1.34   &  0.16   &   2.0   &  2.05   &  254   & 3.1$\times$10$^{-2}$  \nl
 30   &  1.12   &  0.16   &   2.8   &  0.80   &    0   & 1.0$\times$10$^{-5}$  \nl
 31   &  1.15   &  0.07   &   7.4   &  0.84   &    1   & 6.0$\times$10$^{-4}$  \nl
 32   &  1.14   &  0.25   &   4.1   &  1.01   &    0   & 1.5$\times$10$^{-4}$  \nl
 33   &  1.23   &  0.33   &   7.8   &  0.98   &    5   & 1.3$\times$10$^{-3 }$ \nl
 34   &  0.87   &  0.05   &  10.5   &  0.77   &    0   & 1.0$\times$10$^{-5 }$ \nl
 37   &  0.85   &  0.04   &  12.6   &  0.77   &    5   & 1.9$\times$10$^{-3 }$ \nl
 \bf 38a   & \bf  0.97   & \bf  0.02   & \bf   7.6   & \bf  1.31   & \bf    7
 & \bf 1.5$\times$10$^{\bf -3}$  \nl
 38b   &  1.45   &  0.06   &   6.3   &  0.30   &    1   & 1.6$\times$10$^{-3}$  \nl
 39   &  1.15   &  0.02   &   1.6   &  1.75   &   23   & 3.3$\times$10$^{-3}$  \nl
 40   &  1.12   &  0.05   &   2.3   &  1.13   &   13   & 3.1$\times$10$^{-3}$  \nl
 41a   &  0.87   &  0.02   &   2.6   &  0.80   &    5   & 1.9$\times$10$^{-3}$  \nl
 41b   &  1.16   &  0.04   &   5.8   &  0.54   &    5   & 2.8$\times$10$^{-3}$  \nl
\bf 42a   & \bf  0.91   & \bf  0.01   & \bf   4.3   & \bf  0.97   & \bf    3
& \bf 1.0$\times$10$^{\bf -3 }$ \nl
 42b   &  1.40   &  0.09   &   4.2   &  0.61   &   12   & 5.1$\times$10$^{-3}$  \nl
\bf Earth\tablenotemark{d} & \bf 1.00 & \bf 0.03 &  \bf 2.1 &  \bf 1.00 & \bf $\sim$1-10 &  \bf 1.0$\times$10$^{\bf -3 }$ \nl

\enddata
\tablenotetext{a}{Planets with 0.9 AU $<$ a $<$ 1.1 AU are shown in
bold, and depicted in Fig.~\ref{fig:hab}.}
\tablenotetext{b}{Number of Oceans of water accreted by the terrestrial
planet, where an ocean is equal to 1.5 $\times$10$^{24}$ grams of water.}
\tablenotetext{c}{Water Mass Fraction of the planet.}
\tablenotetext{d}{The orbital elements for the Earth are 3 Myr
averages from Quinn et al. (1991).  The water content of Earth's mantle is
uncertain.  We assumed the Earth's total water budget to be four oceans in
calculating the water mass fraction.  See Morbidelli et al. (2000) for a discussion. }
\end{deluxetable}

\normalsize

All but a few of our simulations generate planets within the semi-major axis
space 0.8-1.5 AU (table 2), and these planets acquire a wide range of water
during their buildup from lunar-to-Mars-sized bodies as simulated in this
paper. A useful absolute unit of water content is an "Earth ocean," defined to
be 1.5 x 10$^{24}$ g of water. This is the amount of water in the hydrosphere
(oceans, rivers, lakes, etc.) of the Earth today, excluding an additional 15\%
in the crust of our planet. Highly uncertain is how much additional water is
in the Earth's deep interior today-estimates range from much less than 1 Earth
ocean to several Earth oceans (Abe et al. 2000). The amount present during
accretion of the Earth is even more poorly constrained, because of a lack of
geochemical evidence for the extent of hydration of the primitive mantle. Some
geochemists argue that the Earth possessed large amounts ($>$ 10 Earth oceans)
of water (Dreibus and Wanke 1989), but others assert that the current
inventory is close to what was present at the close of accretion.  

The amount of water accreted by our modeled terrestrial planets, in the
0.9-1.5 AU region, ranges from 0 to almost 300 Earth oceans. These numbers
likely should be halved to account for loss of water during the large embryo
impacts that characterize the growth of the terrestrial planets in our
simulations.  Even so, some planets clearly receive much more water than did
the Earth during its formation, and some ended up essentially dry (those with
a value zero might have some trace amounts of water depending on th water
content of 1 AU planetesimals; see below). There is no correlation in our
simulations between the amount of water acquired and other parameters such as
final planet mass, orbit, or positions and mass of the giant planet in the
simulation. The number of bodies from beyond 2-2.5 AU that end up in the
0.9-1.5 AU region is highly sensitive to the initial conditions, and they
determine the water delivered to the growing terrestrial planets.  The results
reflect well the stochastic nature of the accretion process in a
planet-forming environment stirred up dynamically by Jovian-mass bodies. . 

Some cosmochemists have argued that the planetesimals in the region of 1 AU,
where the Earth formed, were not dry but instead had as much water as was
required to produce what is seen in the Earth today (Drake and Righter
2002). The geochemical arguments regarding a local versus distal source of
water are too involved to get into here (Robert 2001). Were we to assume that
material at 1 AU possesses, let us say for sake of argument, 0.05\% water (2
Earth oceans), then our results would reflect somewhat larger amounts of water
than are shown in table II. But the amount of water in the carbonaceous
chondrites is so large that the shift would not be substantial, except that
the driest planets would have at least 2 Earth oceans rather than 0. The
stochasticity of the results would be preserved, reflecting the chance
delivery of large bodies from the region beyond 2.5 AU that is one of the
signatures of terrestrial planet accretion according to this model.

\subsection{Characterizing the formed terrestrial planets}

We have formed a total of 106 terrestrial planets in 42 simulations.  (Recall
our definition
that a terrestrial planet has $a <$ 2 AU and M $>$ 0.2 $\mearth$.)  This
ensemble is comprised of planets which formed in a wide range 
of environments, and is more suited to examining the possible outcomes of
planet formation rather than the likely outcomes in specific cases or the true
distributions of terrestrial planets in our galaxy.  

\begin{figure}
\psfig{figure=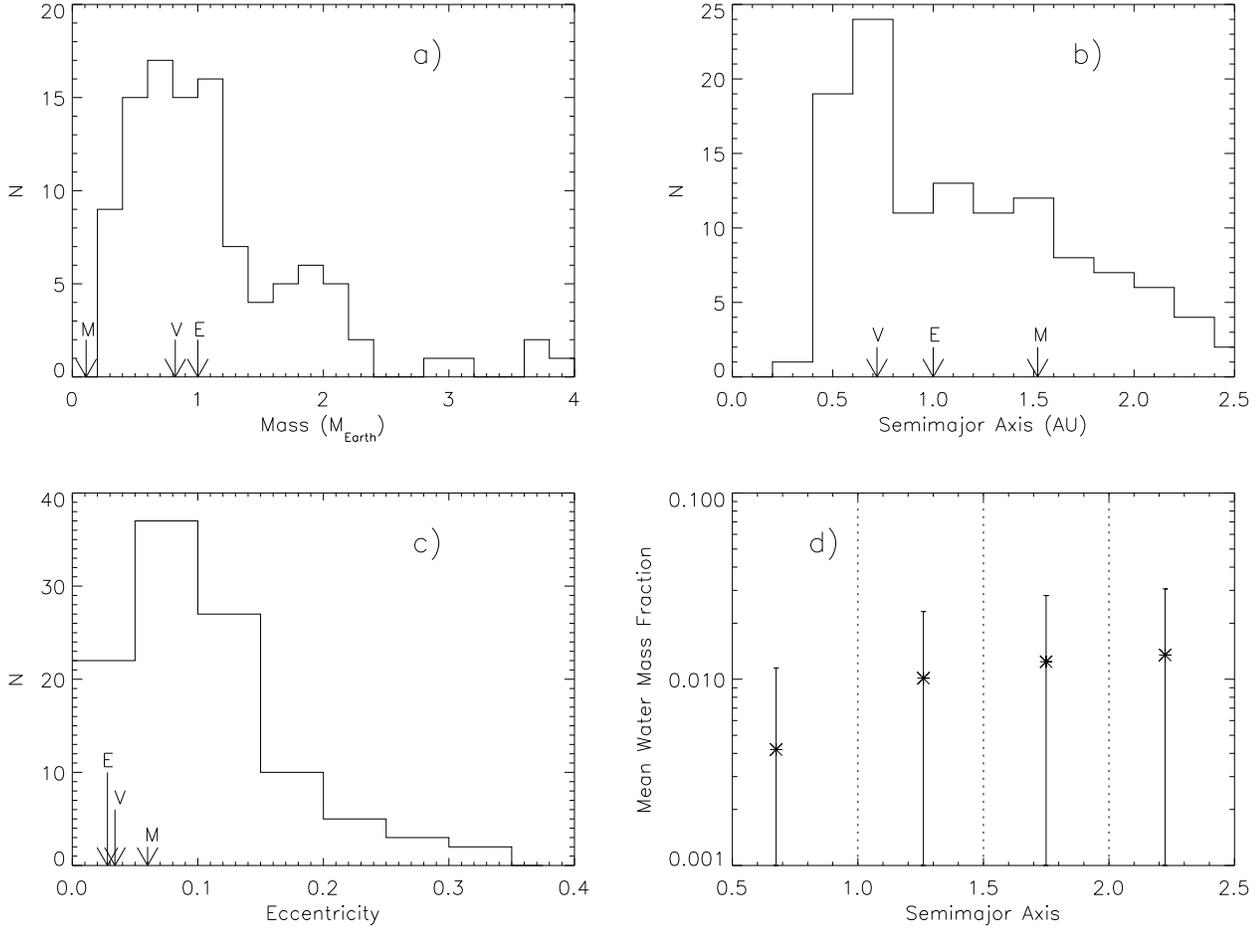,height=13cm,angle=90}
\caption{Orbital and physical characteristics of 106 terrestrial
planets formed in 42 simulations, and 12 planets which formed between 2 and
2.5 AU.  Panels a), b) and c) show mass, eccentricity
and semimajor axis functions.  Values for Venus, Earth and Mars are labeled
with arrows which correspond to their values averaged over a 3 Myr integration
(Quinn et al. 1991).  Panels a) and b) are comprised solely of the planets
inside 2 AU.  Panel d) shows the mean water content of
terrestrial planets as a function of their final semimajor axis, divided into
four zones as indicated by the dotted vertical lines, with one sigma error
bars.  The statistical significance of panels c) and d) is discussed in the text.}
\label{fig:char4}
\end{figure}

Figure~\ref{fig:char4} summarizes the physical properties of these 106
terrestrial planets.  Panels a), b) and c) give the mass, semimajor axis and
eccentricity functions, respectively, of our sample.  Panel a) shows that
one Earth mass is not an upper limit to the mass of terrestrial planets, which
can have masses up to 3-4 $\mearth$.  Panel b) demonstrates that we form
terrestrial planets throughout the terrestrial region, with the innermost
planet forming at 0.34 AU.  Recall, however, that our initial conditions start
with planetary embryos at 0.5 AU due to computational constraints.

Panel c) shows the eccentricity distribution of terrestrial planets (inside 2
AU).  We performed a one-sided Kolmogorov Smirnov test to see if the planets
in our solar system are consistent with having been drawn from this eccentricity function.
If we consider only Earth, Venus and Mars, then the solar system does not
match this distribution to 95\% confidence (90\% if only simulations with
$m_{planetesimal}$ = 0.01 $\mearth$ -- case (ii) above -- are considered).  If
Mercury is included, then the statistic is insignificant, as the solar system
is inconsistent with the distribution with 74\% confidence (53\% for case (ii)
only).  This is not a new problem, as many authors (e.g. Chambers 2001, 2003)
have had trouble matching the low eccentricities and inclinations in our solar
system to numerically formed terrestrial planets, and is, as of yet, unsolved.

Panel d) shows the
radial variation in water content of these planets, divided into four
semimajor axis bins: (1) a $<$ 1 AU, (2) 1 AU $<$ a $<$ 1.5 AU, (3) 1.5 AU $<$ a $<$ 2 AU,
and (4) 2 AU $<$ a $<$ 2.5 AU.  We performed a Wilcoxon test on each two adjacent
bins to test whether the difference in water content of the planets in these
bins is statistically significant.  We found that the difference between bins
1 and 2 was significant to 99.6\%, but that there is no statistical difference
between the means in bins 2, 3 and 4.  If we only include simulations with
$m_{planetesimal}$ = 0.01 $\mearth$ (case (ii) above), the difference between bins 1 and 2 is
still significant to 99.4\%, and there remains no difference between the outer
three bins.

\section{Discussion}

\begin{figure}
\psfig{figure=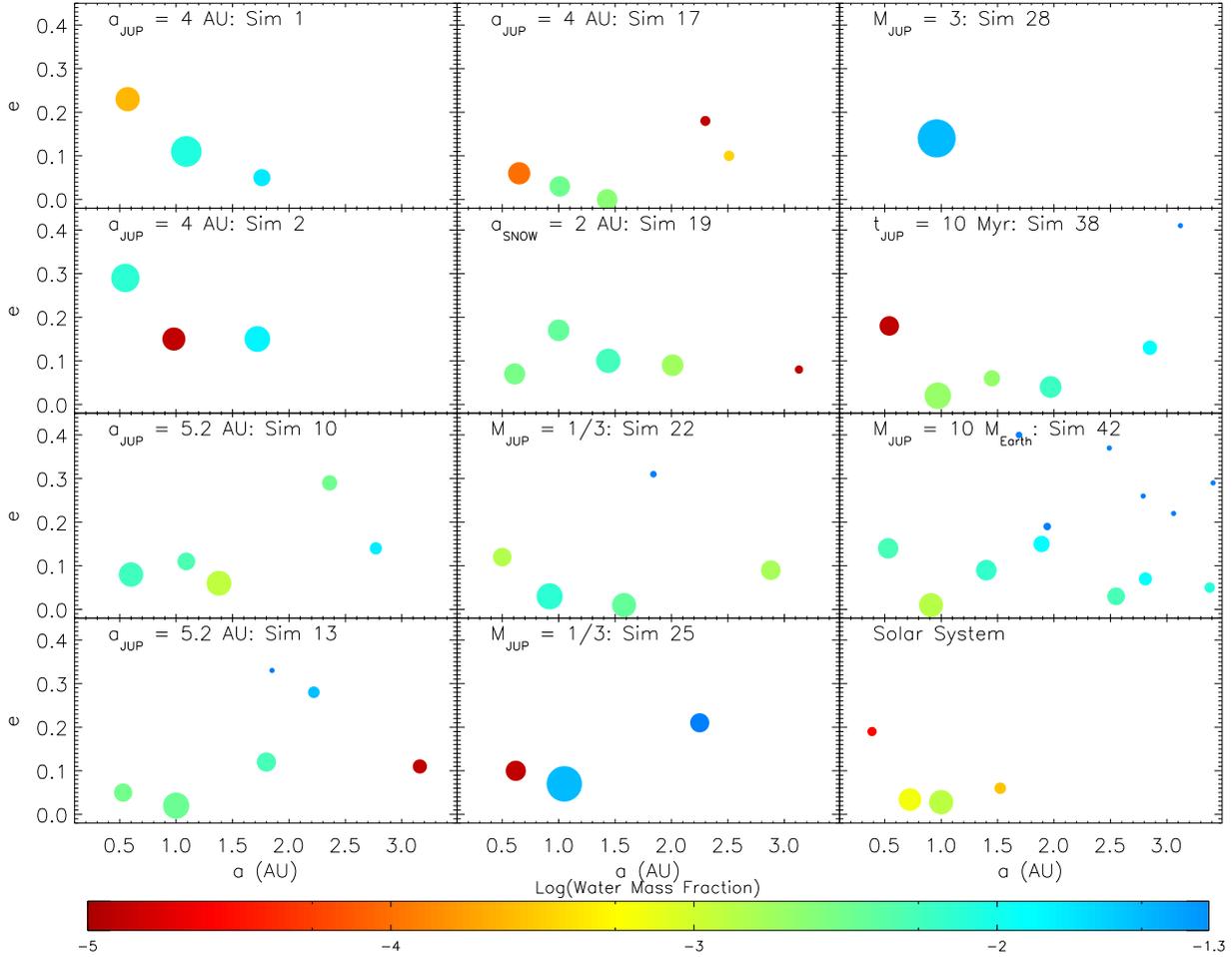,height=13cm,angle=90}
\caption{Final configuration of 11 simulations which formed a ``habitable''
planet with 0.9 AU $<$ a $<$ 1.1 AU, labeled by the physical parameters of
each planetary system and the simulation number.  If not otherwise mentioned,
M$_J$ = M$_{J,r}$ and $e_J$ = 0.  Our solar system is included for comparison,
with 3 Myr averaged values from Quinn et al. (1991). See tables 1 and 2 for
more details.}  
\label{fig:hab}
\end{figure}

In 42 simulations, we have formed terrestrial planetary systems of all shapes
and sizes, with 1-4 terrestrial planets and a range in water content and
orbital characteristics.  The extremes are systems in which only one, very
massive terrestrial planets has formed (e.g. simulation 28: see
Fig.~\ref{fig:hab}) and systems which have many, lower-mass, more closely
packed systems with 4 planets inside 2 AU (e.g. simulation 42: see
Fig.~\ref{fig:hab}).  These results imply the existence of a huge variety of
planetary systems in our galaxy, as planets form from disks
around stars with a variety of masses and metallicities.

Chambers (2003) formed ``life-sustaining'' planets in his simulations, defined to
be in the habitable zone with water mass fractions greater than
4$\times$10$^{-4}$.  He found that the water content of terrestrial planets
depends strongly on the eccentricity, mass and formation time of the giant planets, with
larger values of $e_J$ and M$_J$ leading to drier planets, while larger values
of t$_J$ led to more volatile-rich planets.  He also found that systems with lower
mass giant planets form the most life-sustaining planets.

Our results are partially consistent with Chambers'.  We also find that $e_J$
plays a large role in terms of the water content of terrestrial planets, and
two of the 11 ``habitable'' planets we formed between 0.9 and 1.1 AU were from
only four simulations with M$_J$ = 1/3$M_{J,r}$, and both had enough
water to be deemed ``life-sustaining.''  However, we see no
correlation between Jupiter's mass and the water content of terrestrial
planets.  This could be due to our relatively small number of simulations.  We
have not performed enough simulations with different values of t$_J$ to reach
a reliable conclusion.

\subsection{Applications to TPF/Darwin}

TPF and Darwin are respectively, U.S. and European projects to put large
telescopic systems into space to detect and spectroscopically characterize
Earth-sized planets around other stars, with the goal of identifying those
with spectroscopic signatures suggesting habitability (water) or even life
(molecular oxygen)(Beichman et al. 1999). While efforts are
being made to design the systems to detect Earths in as broad a range of
semi-major axes as possible, detection of an Earth-sized planet in a
Venus-like orbit at 0.7 AU from a solar-type star may pose special problems in
nulling-out or blocking the light of the parent star. However, the results we
show in table 2 suggest that there will be many such candidates in Venus-like
orbits, and hence a complete census of terrestrial planets in
Venus-to-Mars-sized orbits remains a desirable goal. 

Most significant among our results is the robustness of terrestrial planet
formation. This has been shown previously for starting positions of the giant
planets akin to our solar system (Chambers \& Cassen 2002), and we have
extended the results by including giant planets over a wide range of masses
and a range of orbits commensurate with terrestrial planet stability but
varying significantly from the solar system situation. While terrestrial
planet formation---that part of the process in which lunar-to-Mars-sized bodies
are cleaned up to form a few Earth-sized planets---is stochastic, it still leads
to planets familiar to us in terms of size and position. Of course, one can
start with initial conditions so different from our solar system that
predominantly smaller or larger planets result, and indeed if residual nebular
gas is included in the simulation, numerous Mars size bodies replace the
handful of Earth-size objects (Kominami and Ida, 2002). But in the coarse
view, it seems easy to make terrestrial planets and hence the TPF/Darwin
projects ought to assume a high likelihood of terrestrial planets around
solar-type stars for which giant planets are not so close as to induce orbital
instability.  

\begin{figure}
\centerline{\psfig{figure=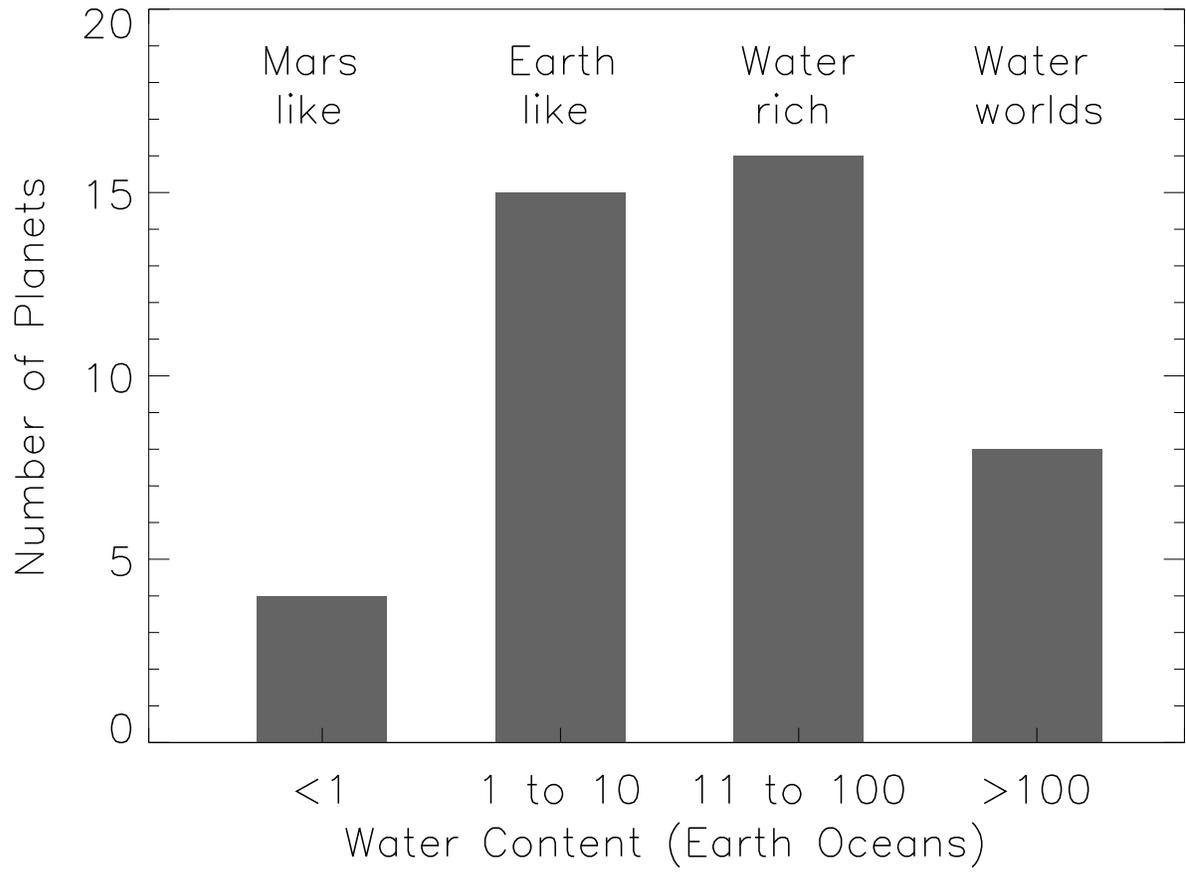,height=13cm}}
\caption{Histogram of the water content of 43 planets with 0.8 AU
$<$ a $<$ 1.5 AU which formed in 42 simulations.  See text for discussion.}
\label{fig:hist}
\end{figure}

Our results also predict, as discussed above, a wide range of possible water
abundances, which we summarize in Fig.~\ref{fig:hist}.
``Mars-like'' worlds have less than an Earth ocean of water (Lunine et al.,
2003).  ``Water-rich'' worlds may be so wet as to have a geologic evolution
different from ours; ``water worlds'' are dominated by deep water mantles and
remain speculative (Leger et al., 2003).
The spectroscopic signature of water vapor in an atmosphere is
such as to be an insensitive diagnostic of water abundance---other than
indicating that indeed surface water is present and surface temperatures warm
enough for significant water vapor. Yet, the bulk water abundance will
certainly affect the evolution of a planet, in terms of tectonic styles (e.g.,
plate tectonics on Earth versus dominant basaltic volcanism on Venus), hence
cycling of volatiles, and other processes that in complex ways determine
habitability. Modeling such planets provides a perspective on these
dependencies, but as yet unknown is how to remotely assess another Earth as
nearly dry or so rich in water as to be a novel, ``water-world'' terrestrial
planet. Our results suggest that both ends of the spectrum must be seriously
considered.  

\section{Conclusions}

We have performed 42 simulations of terrestrial planet formation, with initial
conditions designed to reflect the state of the protoplanetary disk at the 
end of oligarchic growth (see table 1).  These simulations produced a variety of
planetary systems.  Planets formed with masses from 0.23 $\mearth$ to 3.85
$\mearth$, and with water contents ranging from completely dry to ``water
worlds'' with 300+ oceans of water.  A total of 43 planets formed between 0.8
and 1.5 AU, including 11 ``habitable'' planets between 0.9 and 1.1 AU (see
table 2 and Fig.~\ref{fig:hab}).

Terrestrial planet formation is a robust process, and a
stochastic one.  Stochastic ``noise'' between simulations with similar
initial conditions made it difficult to spot trends with certain parameters.
We found that the parameter with the strongest
effect on the terrestrial planets was the planetesimal mass we chose, 
reflecting the surface density past the snow line.  A high density (and large
planetesimal mass) in this region results in the formation of a smaller number
of terrestrial planets with larger masses and higher water content, as compared
with planets which form in systems with lower densities past the snow line (and
smaller planetesimal masses).  

Jupiter's eccentricity plays an important
role in the volatile delivery process, as even a modest eccentricity of 0.1
drastically reduces the water content of the terrestrial planets.  Systems
with $e_J$ $>$ 0 tend to form terrestrial planets with
slightly higher eccentricities than those with $e_J$ = 0, and the total mass
in terrestrial planets is less for systems with eccentric Jupiters.  This is
significant in light of the high eccentricities of discovered extrasolar
planets. 

In the cases of Jupiter at 7 AU, or a snow line of 2 AU and Jupiter at 5.2 or
7 AU, our model predicts the formation of 1-2 $\mearth$ ``super embryos'',
protoplanets which form in a region of enhanced density between the snow line
and 3:1 Jupiter resonance.  These super embryos serve as a small dynamical
barrier for inward-diffusing, volatile-rich planetesimals.  This is reflected
in the very low mean water content of the innermost terrestrial planet in
systems with $a_J$ = 7 AU.  Super embryos can also serve as the accretion seed
for massive terrestrial planets with high water contents.  

We believe our sample to be representative of the extremes of terrestrial
planet formation under our assumed initial conditions (i.e. what is possible),
rather than to be characteristic of the planets in our galaxy.  It is unclear
at the present which initial conditions are the most realistic.  

In future work we intend to improve the resolution of these simulations by
increasing the number of particles by an order of magnitude.  We will probe
new regions in parameter space, in order to further improve our understanding
of how terrestrial and habitable planets form.

\section{Acknowledgments}

JL is grateful to the NASA Planetary Atmospheres program for support.  
SR and TQ are grateful to the NASA Astrobiology Institute for
support.  Many of the simulations presented here were performed on
computers graciously donated by Intel.  SR thanks John Chambers for the use of
his code (Mercury) and insightful discussions, and Don Brownlee for many
constructive conversations.  

\pagebreak

\end{document}